\documentstyle[12pt]{article}

\def\Re{\,\hbox{{\bf Re}}\,}

\def\epm#1#2{\hbox{${\lower1pt\hbox{$\scriptstyle +#1$}}
\atop {\raise1pt\hbox{$\scriptstyle -#2$}}$}}

\def\gsim{\mathrel{\rlap{\lower4pt\hbox{\hskip1pt$\sim$}}
    \raise1pt\hbox{$>$}}}         

\def\frac#1#2{{{#1}\over {#2}}}

\catcode`@=11 
\renewcommand\section{\@startsection {section}{1}{\z@}
    {-3.5ex plus -1ex minus -.2ex}{2.3ex plus .2ex}{\bf}}
\renewcommand\subsection{\@startsection {subsection}{1}{\z@}
    {-3.5ex plus -1ex minus -.2ex}{2.3ex plus .2ex}{\it}}
\def\slash#1{\mathord{\mathpalette\c@ncel#1}}
 \def\c@ncel#1#2{\ooalign{$\hfil#1\mkern1mu/\hfil$\crcr$#1#2$}}
\def\lsim{\mathrel{\mathpalette\@versim<}}
\def\gsim{\mathrel{\mathpalette\@versim>}}
 \def\@versim#1#2{\lower0.2ex\vbox{\baselineskip\z@skip\lineskip\z@skip
       \lineskiplimit\z@\ialign{$\m@th#1\hfil##$\crcr#2\crcr\sim\crcr}}}
\catcode`@=12 

\newcommand{\eq}[1]{Eq.~(\ref{#1})}

\newcommand{\nl}{\nonumber \\}

\def\beq{\begin{equation}}
\def\eeq{\end{equation}}
\def\beqa{\begin{eqnarray}}
\def\eeqa{\end{eqnarray}}

\begin{document}
\begin{titlepage}
\setcounter{page}{0}
\begin{flushright}
{\tt hep-ph/9812479}\\
{DFTT 74/98}\\
{INFN-RM3 98/9}\\
\end{flushright}

\vskip.2cm
\begin{center}
{\Large \bf Truncated moments of parton distributions} \\

\vskip 1cm

{\bf Stefano Forte}\footnote{On leave from INFN, Sezione
di Torino, Italy}\\
{\sl INFN, Sezione di Roma III}\\
{\sl Via della Vasca Navale 84, I-00146, Roma, Italy}\\

\vskip .4cm

{\bf Lorenzo Magnea}\\
{\sl Dipartimento di Fisica Teorica, Universit\`a di Torino}\\
 and {\sl  INFN, Sezione di Torino}\\
{\sl Via P. Giuria 1, I-10125 Torino, Italy}\\

\vskip 1.6cm
\end{center}
\begin{abstract}

We derive evolution equations satisfied by moments of parton
distributions when the integration over the Bjorken variable 
is restricted to a subset $(x_0 \le x \le 1)$ of the allowed kinematical 
range $0\le x\le 1$. The corresponding anomalous dimensions turn out to be
given by a triangular matrix which couples the $N$--th truncated moment with 
all $(N + K)$--th truncated moments with integer $K \ge 0$. We show that the 
series of couplings to higher moments is convergent and can be truncated to 
low orders while retaining excellent accuracy. We give an example of 
application to the determination of $\alpha_s$ from scaling violations.

\end{abstract}
\vfill
\leftline{December, 1998}

\end{titlepage}


The description of scaling violations of deep--inelastic
structure functions is historically one of the first predictions of
perturbative QCD amenable to experimental testing, and still is
one of the most accurate ways to use perturbative QCD for
precision measurements~\cite{phrev}. 
Specifically, comparison of the theoretical prediction with the data 
allows a determination of the only free parameter in the QCD lagrangian,
the strong coupling $\alpha_s$, as well as the extraction of the parton
distributions of hadrons which, though in principle computable,
are determined from the nonperturbative dynamics of the theory and
thus must be treated as unknown phenomenological parameters.

As well known, scaling violations are described by ordinary
linear differential equations for evolution in $t\equiv\ln{Q^2\over
\Lambda^2}$ satisfied by the Mellin moments of parton distributions, 
which can be directly viewed as matrix elements of local operators. 
At the leading log level, parton distributions can be expressed directly
as linear combinations of measurable structure functions; beyond leading 
log this is only true in specific factorization schemes, whereas in a 
general scheme the moments of structure functions are related to moments 
of parton distributions through Wilson coefficients which are calculable 
as perturbative expansions in $\alpha_s$~\cite{threv}.

{}From a phenomenological point of view, however, dealing with moments of
structure functions is rather inconvenient, since by definition
Mellin moments are obtained integrating over all values of the Bjorken
variable $0\le x\le 1$. Since $x$ is related to the invariant energy
$W^2$ of the virtual photon-hadron scattering process by $W^2={1-x\over
x}$,  $x\to 0$ is the infinite energy limit and can thus never be attained 
experimentally. All moments are thus subject to an {\it a  priori} infinite 
uncertainty from this region, which can only be reduced on the basis of 
nonperturbative models and assumptions, such as the idea that the virtual 
photon-hadron scattering cross section should behave at high energy like 
the cross section for scattering between real hadrons and thus 
be controlled by Regge theory~\cite{regge}. It follows that any use of the 
evolution equations for moments requires some model-dependent input.

A way to avoid this problem is of course well known: work in $x$-space and
deal with Altarelli-Parisi equations, which give
directly the evolution of parton distributions (rather than of their
moments). Undoing the moments turns the ordinary differential
equations in $t$ into integro-differential equations in $x$ and $t$,
but the $x$ integration is such that scaling violations at $x_0$ only
depend on the values of parton distributions for all $x\ge
x_0$. The need for an extrapolation to the unmeasurable
$x\to0$ region is thus at least in principle avoided~\cite{ap}.

In practice, however, in order to be able to solve the Altarelli-Parisi
equation, we must parameterize the data in some specific way. This is usually 
done by fitting a well-defined functional form $f(x)$, parametrized by some
free parameters, to data at a fixed scale $Q^2$; or, alternatively, expanding 
on suitable basis of functions~\cite{pet}.
In either case, a bias is introduced in the analysis, due to the
fact that the specific choice of functional form of the fitting 
function, or of the basis functions, constrains, for obvious reasons of 
smoothness, the description of data with the smallest measured values of $x$.
Giving a precise quantitative estimate of the possible bias thus
introduced can in practice be rather complicated~\cite{abfr}.

One may also be interested in using scaling violations for the determination 
of a moment of a parton distribution. For instance, the gluon distribution
is primarily determined from scaling violations~\cite{cteqglu}, and one may 
be interested directly in its moments because of their physical interpretation,
or because the structure of evolution equations is such that some moments
are better constrained by an observed pattern of scaling violations than the 
parton distribution for any individual value of $x$~\cite{posmon}. However, 
when arriving at a determination of, say, the first moment of the polarized
gluon distribution~\cite{abfr}, it would be very useful to be able to 
separate the  contribution to the moment from the measured region --- which 
is determined from the data up to conventional uncertainties --- from that 
due to an extrapolation which is necessarily entirely based on theoretical
prejudice.

All these problems are solved if one is able to formulate
evolution equations directly in terms of moments restricted to the
measured region. Since, as already mentioned, evolution at $x_0$
requires knowledge of  the parton distribution for all $x \ge x_0$,
it is clear that the scale dependence of a moment evaluated by integrating 
over the restricted range $x_0 \le x \le 1$ (truncated moment, henceforth)
rather than over the full kinematically allowed region $0 \le x \le 1$, is
determined by the structure function in the same
region.\footnote{Turning the argument around, this also means that
evolution equations for moments truncated to a generic interval
$x_0 \le x \le x_1$ do not close, because they  would also require
knowledge of parton distributions for $x > x_1$. Hence, problems related
to the extrapolation to $x = 1$ of data taken only at $x \le x_1$ cannot
be solved by the methods of this paper. The extrapolation to $x = 1$ is
however a much less serious problem, both because the $x \to 1$ limit
is, unlike the $x\to 0$ limit, at least in principle experimentally
accessible, and because structure functions must vanish as $x \to 1$
for kinematic reasons, so that the extrapolation is under much better
control.}    
It is however not obvious that the evolution equations for truncated moments 
will take a simple form, or even that they will close upon a specific
subset of moments: if they did not, nothing would be gained by introducing
truncated moments over the resolution of the full Altarelli-Parisi
equations. 

Here, we will derive evolution equations for truncated moments.
We will show that the evolution of truncated moments is driven by a 
triangular matrix of anomalous dimensions which couples the $N$-th truncated 
moment to all $N+K$-th moments, where $K$ is a positive integer (but $N$ 
can be any real number).  The elements of this matrix of anomalous dimensions 
depend on the cutoff $x_0$, and are calculable in perturbation theory as 
straightforward integrals of the  Altarelli-Parisi splitting functions.
Furthermore, we will prove that the series of couplings to higher moments is 
convergent, so that the infinite matrix of anomalous
dimensions can be truncated to specified accuracy. We will also see
that this convergence is very fast, so in practice it is only necessary to 
deal with small (typically less than $10\times10$) matrices.

In order to derive the evolution equations for truncated moments, we
start from  the Altarelli-Parisi evolution equation 
\beq
\frac{d}{dt}~q(x, Q^2) =
\frac{\alpha_s (Q^2)}{2 \pi} \int_x^1 \frac{d y}{y} 
P\left(\frac{x}{y};\alpha_s(Q^2)\right) q(y, Q^2)~~.
\label{alpar}
\eeq
Here $q(x, Q^2)$ is the nonsinglet quark distribution; in the
singlet case one must consider a $2\times2$ matrix of anomalous
dimensions which mixes the quark and gluon distributions. This
introduces some trivial complications which we will not discuss here.
The splitting function $P(x)$ is given as a series in $\alpha_s$,
$P(x)=P^{(0)}+{\alpha_s\over 2\pi}P^{(1)}+\dots$;
henceforth we will suppress its explicit $\alpha_s$ 
dependence.
 
The truncated Mellin moment of the parton distribution $q(x, Q^2)$ is 
defined as
\beq
q_N(x_0, Q^2) \equiv \int_{x_0}^1 d x x^{N - 1} q(x, Q^2)~~.
\label{trunc}
\eeq
By integrating \eq{alpar} and inverting the order of the double
integration it is easy to see that truncated moments satisfy the evolution
equation
\beq
\frac{d}{dt} q_N(x_0, Q^2) =
\frac{\alpha_s (Q^2)}{2 \pi} \int_{x_0}^1 d y y^{N - 1} q(y, Q^2) 
G_N\left(\frac{x_0}{y};\alpha_s(Q^2)\right)~~,
\label{truncalpar}
\eeq
with an evolution kernel $G_N$ given by a truncated moment of the
splitting function $P(z)$,
\beq
G_N(x) = \int_x^1 d z z^{N - 1} P(z)~~.
\label{kern}
\eeq

If $x_0 = 0$ the kernel $G_N(x_0/y)$ reduces to the usual $x$-independent
anomalous dimension, $G_N(0) = \gamma_N$, it can thus be taken outside
the integral in \eq{truncalpar}, and the {\it r.h.s.} of 
\eq{truncalpar} depends only on the $N$-th moment. Henceforth we 
will assume that $\Re N$ is large enough for $\gamma_N$ to be regular.
If instead $x_0 \not= 0$, because of the residual $y$ dependence in the 
kernel $G_N$ the evolution equation does not diagonalize.
However, it is easy to see that the $N$-th truncated moment mixes only 
with moments with index $M \geq N$. 

To prove this, expand the kernel $G_N$ in Taylor series around $y = 1$.
\beq
G_N \left(\frac{x_0}{y}\right) = \sum_{n=0}^\infty
\frac{g_n^N (x_0)}{n!} (y - 1)^n~~,
\label{ser1}
\eeq
where
\beq
g_n^N (x_0) \equiv \frac{\partial^n}{\partial y^n} 
	G_N \left. \left(\frac{x_0}{y}\right) \right|_{y = 1}~~.
\label{co1}
\eeq
Since $G_N(z)$ is regular for all $0 \le z < 1$, but in general has
logarithmic singularities as $z \to 1$, due to the presence of $+$ 
distributions in the splitting function $P(z)$, the Taylor expansion in 
\eq{ser1} has radius of convergence $r = (1 - x_0)$. However, 
since the singularities of $G_N(x_0/y)$ at $y = x_0$ are integrable,
we can substitute the Taylor expansion \eq{ser1} in the {\it r.h.s.}
of the evolution \eq{truncalpar}, exchange the order of sum
and integral, and still end up with a convergent sum.  Since
$G_N(x_0/y)$ is regular at $y = 1$, the 
expansion \eq{ser1} contains only non--negative powers of $y$, so,
after substitution of the expansion in \eq{truncalpar}, there is 
no mixing between the $N$-th truncated moment and moments with $M < N$, as 
promised. 

Because of the convergence of the Taylor expansion (\ref{ser1})
and of the ensuing expansion of the {\it r.h.s.} of \eq{truncalpar},
we can truncate the expansion at
finite order $M$. A straightforward computation then leads to
\beq
G_N \left(\frac{x_0}{y}\right) = \sum_{K=0}^M c_{K,N}^{(M)} (x_0) y^K +
O\left[(y-1)^{M + 1}\right]~~,
\label{ser2}
\eeq
where
\beq
c_{K,N}^{(M)}(x_0) = \sum_{p=K}^M 
\frac{(-1)^{K + p} g_p^N (x_0)}{K! (p - K)!}~~,
\label{co2}
\eeq
so that the evolution equation (\ref{truncalpar}) becomes
\beq
\frac{d}{dt} q_N(x_0, Q^2) =
\frac{\alpha_s (Q^2)}{2 \pi} \sum_{K=0}^M c_{K,N}^{(M)}(x_0) 
q_{N + K}(x_0, Q^2)~~.
\label{finsyst}
\eeq

We are thus led to  an ordinary finite system of differential equations, which 
can be solved by standard methods, provided the number of equations in the 
system equals the number of unknowns. For this to happen, we must include 
a decreasing number of terms in the series (\ref{ser1}) as the order of the 
moment increases. For example, if one is interested in the evolution of the 
$N_0$--th moment, and wishes to include $M + 1$ terms in the series that 
expresses its scale dependence, one must include in the series associated 
with the $(N_0 + K)$--th moment only  $M + 1 - K$ terms. 
This then gives an upper triangular matrix of coefficients. 
Such an approximation is only possible if higher moments have a decreasing 
influence on the evolution, so they may be approximated less accurately. 

It is easy to see that this is indeed the case. If the Taylor expansion 
\eq{ser1} is truncated at order $M$, the percentage error on the {\it r.h.s.} 
of the evolution \eq{finsyst}, due to the truncation of the series, is equal 
to  
\beq
R(N,M;x_0,Q^2) \equiv \frac{1}{{\cal N}}
\int_{x_0}^1 d y y^{N - 1} q(y, Q^2) \left[
G_N \left(\frac{x_0}{y}\right)-\sum_{K=0}^M c_{K,N}^{(M)}(x_0) y^K \right]~~,
\label{errint}
\eeq
where the normalization is given by the exact integral
\beq
{\cal N} = \int_{x_0}^1 d y y^{N - 1} q(y, Q^2) 
G_N \left(\frac{x_0}{y}\right)~~.
\label{normint}
\eeq
Now, the coupling of the $N$--th moment to the $(N + M)$--th moment
is due to terms which are at least of order $M$ in the Taylor
expansion \eq{ser1} (because the expansion of the coefficient
$c_{K,N}$ in \eq{co2} starts at order $K$). Such terms decrease very rapidly
because the Taylor series is convergent, and furthermore the relative size 
of the $(N + M)$--th moment compared to the $N$--th moment decreases rapidly,
since parton distributions fall off  as a power of $(1 - x)$ as $x \to 1$. 
We will check this  explicitly below in the particular case of the NLO
evolution of the nonsinglet quark distribution, however we emphasize that
it follows from rather general properties 
of parton distributions and their evolution.

Once the system (\ref{finsyst}) has been truncated to a finite size,
it is straightforward to solve the evolution equations explicitly,
by techniques analogous to those used to solve the standard
coupled singlet evolution equations.
At leading order, the evolution kernels are $t$--independent,
so the only $t$ dependence on the {\it r.h.s.} of the evolution equation
is in the explicit factor of $\alpha_s (Q^2)$. The solution is
thus simply obtained  by diagonalizing the matrix of coefficients, a task 
which is in our case is enormously simplified by the fact that the matrix 
is upper triangular.

Let us derive the leading order solution to the evolution equation
of the $N_0$--th truncated moment, using \eq{finsyst}, which we can rewrite
in simplified notation as
\beq
\frac{d q_K}{d t} = \frac{\alpha_s (Q^2)}{2 \pi}
\sum_{L = N_0}^{N_0 + M} C_{K L} q_L~~,
\label{simpLO}
\eeq
where $N_0 \leq K,L \leq N_0 + M$, and the matrix of coefficients is given by
\beq
\left\{
\begin{array}{cccc}
C_{K L} & = & c_{L - K,K}^{(M - K + N_0)} (x_0) & (L \geq K)~~, \\
C_{K L} & = & 0 &(L < K)~~.
\end{array}
\right.
\label{matr0}
\eeq
A few basic properties of triangular matrices are collected in the Appendix.
One of them, useful below, is the fact that the eigenvalues of a triangular 
matrix such as $C$ coincide with the diagonal elements, $C_{K K} = 
c_{0,K}^{(M - K + N_0)}$.

Define now the rotated moments, in the basis in which $C$ is diagonal 
\beq
\hat{q}_K = \sum_{L = N_0}^{N_0 + M} R_{K L} q_L~~,
\label{rot0}
\eeq
where
\beq
\sum_{L,P = N_0}^{N_0 + M} R_{K L} C_{L P} R^{-1}_{P Q} =
C_{K K} \delta_{K Q}~~.
\label{diag0}
\eeq
The matrix $R$ which diagonalizes $C$, and its inverse$R^{-1}$, are also 
upper triangular, and can both be computed exactly by means of a simple 
recursion relation in terms of the elements of $C$, without having to resort 
to the time-consuming evaluation of determinants (see the Appendix).
It is apparent that the rotated moments evolve independently, and
the solution to their evolution equation is given by the familiar expression
\beq
\hat{q}_K (x_0,Q^2) =
\left[\frac{\alpha_s(Q_0^2)}{\alpha_s(Q^2)}\right]^{C_{K K}/b_0}
{\hat{q}_K (x_0,Q_0^2)}~~,
\label{LOsol}
\eeq
where $b_0$ is the leading coefficient of the $\beta$ function, $b_0 = 
11/2 - n_f/3$ for $SU(3)$.
The solution to the evolution equation for truncated moments is then
found by simply rotating back,
\beq
q_K = \sum_{L = N_0}^{N_0 + M} R^{-1}_{K L} \hat{q}_L~~.
\label{rotback}
\eeq

At next-to-leading order, the evolution kernel for truncated moments, $G_N$,
acquires a scale dependence though $\alpha_s(Q^2)$, and can be written as
\beq
G_N \left(\frac{x_0}{y},\alpha_s(Q^2)\right) =
G_N^{(0)} \left(\frac{x_0}{y}\right) + \frac{\alpha_s(Q^2)}{2 \pi}
G_N^{(1)} \left(\frac{x_0}{y}\right)~~.
\label{nlogn}
\eeq
The NLO evolution equation can then be written as
\beq
\frac{d q_K}{d t} = \frac{\alpha_s (Q^2)}{2 \pi}
\sum_{L = N_0}^{N_0 + M} \left[ C^{(0)}_{K L} +
\frac{\alpha_s (Q^2)}{2 \pi} C^{(1)}_{K L} \right] q_L~~,
\label{simpNLO}
\eeq
where the matrices $C^{(0)}$ and $C^{(1)}$ are given by \eq{matr0}
in terms of the coefficients $c_{K,N}^{(0)}$ and $c_{k,N}^{(1)}$ constructed 
according to Eqs.~(\ref{ser1}-\ref{co2}) from the LO and NLO kernels 
respectively.

Diagonalizing the matrix of leading order anomalous dimensions with the
matrix $R$ of Eqs. (\ref{rot0}) and (\ref{diag0}) we get an evolution 
equation for the rotated moments $\hat{q}$, of the form
\beq
\frac{d \hat{q}_K}{d t} = \frac{\alpha_s (Q^2)}{2 \pi}
\sum_{L = N_0}^{N_0 + M} \left[ C_{K K} \delta_{K L} +
\frac{\alpha_s (Q^2)}{2 \pi} \hat{D}_{K L} \right] \hat{q}_L~~,
\label{diagNLO}
\eeq
where
\beq
\hat{D}_{K L} = \sum_{P,Q = N_0}^{N_0 + M} R_{K P}
C^{(1)}_{P Q} R^{-1}_{Q L}~~.
\label{nlorot}
\eeq
The matrix evolution equation (\ref{diagNLO}) can be solved with standard
techniques of perturbation theory. The evolved (rotated) moments are expressed
in terms of the initial condition and of the various anomalous dimensions 
involved as
\beqa
\hat{q}_K (Q^2) & = & \left[\frac{\alpha_s(Q_0^2)}{\alpha_s(Q^2)}
	\right]^{C_{K K}/b_0} \left[ 1 - \frac{C_{K K} b_1}{2 \pi b_0^2} \left(
	\alpha_s(Q_0^2) - \alpha_s(Q^2) \right) \right] 
	\hat{q}_K (Q_0^2) \nl
&  - &  \sum_{L = N_0}^{N_0 + M} \frac{\hat{D}_{K L}}{2 \pi}
	\frac{1}{C_{K K} - C_{L L} + b_0} \left[
	\left(\frac{\alpha_s(Q_0^2)}{\alpha_s(Q^2)}\right)^{C_{L L}/b_0} 
	\alpha_s(Q^2) \right. \\ 
& &     \left. - \left(\frac{\alpha_s(Q_0^2)}{\alpha_s(Q^2)}
	\right)^{C_{K K}/b_0} 
	\alpha_s(Q_0^2) \right] \hat{q}_L (Q_0^2)~~, \nonumber
\label{master}
\eeqa
where $b_1$ is the second coefficient of the QCD $\beta$ function,
$b_1 = 51/2 - 19 n_f/6$ for $SU(3)$.
The NLO solution is then found once again by just rotating back
according to \eq{rotback}.

In order to show this formalism at work, we consider now the
determination of $\alpha_s$ from scaling violations of the nonsinglet
structure function $F_2^{\rm NS}$.
Our method allows a direct determination of the strong coupling
by fitting the evolution of the truncated moments of the measured
structure function, with $\alpha_s$ left as the only free parameter,
without having to introduce a parametrization of parton
distributions. A determination of $\alpha_s$ from scaling violations
of a nonsinglet moment has been sometimes attempted, but only in the
presence of sum rules which fix the asymptotic normalization of the
moment~\cite{abfr,ccfr}. However, even the presence of this constraint 
does not obviate the problem of the uncertainty introduced by the small $x$
extrapolation, which remains sizable~\cite{phrev,abfr}. 

The  structure function  $F_2^{\rm NS}$ in the DIS scheme~\cite{dis} is 
simply equal to the nonsinglet quark distribution,
\beqa
\label{f2ns}
F_2^{\rm NS}(x,Q^2)&\equiv&\left(F_2^{\rm p}(x,Q^2)-
F_2^{\rm n}(x,Q^2)\right)\nonumber\\
&=&\sum_{i=1}^{n_f} e^2_i\left[q_i(x,Q^2)+\bar q_i(x,Q^2)\right]\bigg|_{p-n}
\eeqa
In order to determine its evolution it is thus sufficient to transform
to the DIS scheme the well-known NLO nonsinglet splitting function
$P_{qq}^{\rm NS}$~\cite{cfp}. We can then determine the evolution kernel 
$G_N(z)$ \eq{kern} analytically, and study the accuracy of the truncation 
of the expansion in \eq{ser1} by explicitly computing the function 
$R(N,M;x_0,Q^2)$ defined in \eq{errint}. For this purpose, we must use an 
explicit nonsinglet quark distribution, which we can take from any recent 
parton distribution set. We then evaluate the function $R$ at the reference 
scale of the chosen parton set, with several typical choices of  
the lower limit $x_0$ of the $x$--range, by including a decreasing
number of terms as the order of the moment increases, as discussed above. 

The results are shown in Table~1 for moments between the second and the fifth.
It is apparent that, despite the fact that less terms are included, the 
accuracy of the determination of higher moments is actually higher: the fact 
that higher moments are largely insensitive to the lower limit of integration 
overwhelms the error introduced by the truncation of the Taylor expansion in 
\eq{ser1}. In particular, it is apparent that in order to reliably compute the
evolution of the second  moment it suffices to consider a four by four 
evolution matrix. An accurate description of the scaling violations of the
first moment would instead require the inclusion of several more terms; this 
is a consequence of the fact that the integrand in \eq{kern} decreases as 
$z \to 0$ only if $N>1$. The convergence to the correct result is of
course slower when  $x_0$ is larger, since the dependence on $y$
of the kernel \eq{ser1} is weaker when $x_0$ is smaller (and indeed,
there would be none in the limit $x_0 = 0$).
\begin{table}
\begin{center}
\begin{tabular} {|l||l|l|l||l|l|l|}\hline
$x_0$ &$0.01$ &$0.03$ &$0.1$
 &$0.01$ &$0.03$ &$0.1$\\
\hline\cline{1-1}
N   &\multicolumn{3}{c||}{LO}&\multicolumn{3}{c|}{NLO}\\
\hline\hline
2 &$6.3\,10^{-3}$&$3.3\,10^{-2}$&$1.5\,10^{-1}$&$3.5\,10^{-3}$&$2.7\,10^{-2}$&$2.0\,10^{-1}$ \\
\hline
3 &$1.0\,10^{-4}$&$1.7\,10^{-3}$&$3.0\,10^{-2}$&$6.3\,10^{-5}$&$2.8\,10^{-3}$&$3.3\,10^{-2}$ \\
\hline
4 &$1.7\,10^{-6}$&$8.6\,10^{-5}$&$5.1\,10^{-3}$&$1.1\,10^{-6}$&$6.9\,10^{-5}$&$5.5\,10^{-3}$ \\
\hline
5 &$2.7\,10^{-8}$&$4.1\,10^{-6}$&$8.3\,10^{-4}$&$1.8\,10^{-8}$&$3.3\,10^{-6}$&$8.7\,10^{-4}$ \\
\hline
\end{tabular}
\end{center}

\caption{The  percentage error function $R(N,M;x_0,Q^2)$ defined 
in \eq{errint}, computed from the LO and NLO contributions to the 
nonsinglet splitting function in the DIS scheme, with $M = 5 - N$, the 
values of $N$ and $x_0$ shown, $Q^2 = 2.56$~GeV$^2$ and nonsinglet quark 
distribution from the CTEQ4D parton set~\cite{cteq}.}
\end{table}

The simplest way to determine $\alpha_s$ is to
extract from the data the required set of moments, for instance the 
moments from the second to the fifth, as in Tab.~1, at the scale where 
the kinematic coverage in $x$ is widest at large $x$ (so all moments can 
be reliably determined), then solve the evolution equation, compare to 
the data for the second moment at other scales (where, for example, the 
coverage at large $x$ might be smaller so higher moments are less accurately 
determined), and perform a fit of $\alpha_s$.
Our purpose here is not to perform a detailed phenomenological analisys, but 
rather to explore the viability of the method. 

We have thus simply attempted 
such a fit of $\alpha_s$ by using as ``data'' a parametrization of all 
available data on $F_2(x,Q^2)$ for proton and deuteron targets~\cite{tul}, 
which (neglecting nuclear effects) determines the nonsinglet as $F_2^{\rm
NS}=2(F_2^{\rm p}-F_2^{\rm d})$; the moments are then simply found by
numerical integration of the parametrization. 
The kinematic range is essentially limited by the availability of deuterium 
data: even with a truncation point  $x_0=0.1$, a reliable reconstruction of 
the moments is possible only for $30$~GeV$^2\lsim Q^2\lsim100$~GeV$^2$. 
Imposing that power corrections be negligible requires the lower cut at 
$Q^2>30$~GeV$^2$. In fact, since power corrections are large at large
$x$~\cite{phrev}, this is important in order for the determination of the
higher moments to be reliable. 

With all these cuts, fitting to such 
``data'' gives $\alpha_s(M_z)=0.115$. An estimate of the statistical error 
is unfortunately impossible, since a reliable determination of the
covariance matrix for the best-fit parameters is not available~\cite{priv}  
for the fits of Ref.~\cite{tul}.\footnote{Since the parametrization
of $F_2$~\cite{tul} is provided with an ``estimated error band'' one might 
hope to get a qualitative idea of the error by taking the integrals of
the upper and lower curves of the band as estimates of the error on
the moment. This procedure is however meaningless, as seen by noting that 
the error on $\alpha_s$ could then be made arbitrarily small by increasing 
the number of values of $Q^2$ at which the moment is evaluated (even within 
a fixed range in $Q^2$). This apparently paradoxical result is of course due 
to the fact that the procedure neglects correlations between the values of
the moment extracted from the fit at two different scales, which tend to one 
as the scales get closer.}
However, the fact that the central value is so close to the current
global DIS average~\cite{pdg} [$\alpha_s(M_z) = 0.117 \pm 0.002({\rm
exp.}) \pm 0.004({\rm th.})$] suggests that the statistical error is
rather small. 

It thus appears worth considering an actual extraction of $\alpha_s$ from 
the data using this method, either by computing the moments numerically from 
a single set of data, or by first determining a global
parametrization of the data with correlations taken into account as
required. Such an extraction could presumably be improved by
optimizing the choice of moment to be fitted.
Indeed, high moments cannot be determined accurately from
the data, due to the poor knowledge of structure functions at large $x$,
while the evolution of the first moment (which is the lowest convergent 
one in the nonsinglet case) is hard to determine accurately due to the slower
convergence of the expansion \eq{ser1}. Note that the optimal moment
need not be integer, so one would rather expect to have an optimal
range. Determinations of $\alpha_s$ from different moments in this range
could then be combined by properly taking their correlations
into account. 
The extension of the formalism to the singlet sector (which is
essentially straightforward) will allow both a determination
of $\alpha_s$ from wider data sets, and a determination of the partial
moments of the gluon distribution, which, as we already discussed,
might be of great phenomenological relevance.

In summary, we have determined the evolution equations for truncated
moments of parton distribution, and given their explicit
solution to NLO in the nonsinglet sector. From a theoretical
viewpoint, this fills an obvious gap in the available abundant literature on
QCD evolution equations and the methods for solving them. 
{}From a phenomenological viewpoint, our results are a useful addition
to the set of tools available to extract information from the data on
scaling violations, and in particular provide a new way of dealing
with the well-known problem of working around our ignorance of the
small--$x$ behavior of structure functions.
A preliminary extraction of $\alpha_s(M_z)$ from the scaling 
violations of the nonsinglet second moment looks very promising. While we 
postpone to future work a fuller phenomenological analysis, we encourage 
experimental collaborations, which necessarily have a much better control 
of the experimental systematics, to use the simple technique presented in 
this paper as a means to present data on the moments of the gluon 
distribution, as well as to obtain determinations of $\alpha_s$, which would 
be significantly less dependent on model assumptions, in comparison to those 
obtained with more standard techniques.

\vskip 2cm

\noindent{\Large {\bf Appendix}}

\vskip 1cm

We list here a few useful properties of triangular matrices.
Consider a generic $n \times n$ upper triangular matrix $T_n$, with matrix 
elements $a_{i j}$, ($a_{i j} = 0$ for $i > j$).
It is straightforward to show that:

\begin{itemize}
\item[a)] The matrices $T_n$ form a proper subgroup of $GL(n)$.
\item[b)] The eigenvalues of $T_n$ coincide with the diagonal entries
$a_{j j}$. This can be seen noting that the secular equation receives 
contributions only from the diagonal, since all other minors
of $T_n - \lambda {\bf 1}$ vanish.
\item[c)] Let $E_n$ be the matrix of right eigenvectors of $T_n$, arranged 
in columns. $E_n$ is also an upper triangular matrix,
which can be chosen to have all diagonal elements equal to unity. 
Specifically, let ${\bf \beta}^{(j)}$ be the $j$-th right eigenvector of $T_n$,
defined by
\beq
T_n {\bf \beta}^{(j)} = a_{j j} {\bf \beta}^{(j)}~~,
\label{a1}
\eeq
and let $\beta_i^{(j)}$ be the $i$-th component of the $j$-th eigenvector.
Then one can choose $\beta_j^{(j)} = 1$, and one sees that 
$\beta_i^{(j)} = 0$ for $i > j$. Furthermore, the matrix $E_n$, with elements
$E_{n, i j} = \beta_i^{(j)}$, satisfies
\beq
E_n^{-1} T_n E_n = {\rm diag}\left(a_{j j}\right)~~.
\label{a2}
\eeq
\item[d)] The nonvanishing elements of $E_n$ satisfy the recursion relation
\beq
\beta_i^{(j)} = \frac{1}{a_{j j} - a_{i i}} 
	\sum_{p = i + 1}^j a_{i p} \beta_p^{(j)}~~,
\label{a3}
\eeq
initialized by $\beta_j^{(j)} = 1$. This recursion relation can actually 
be solved explicitly in terms of minors of the matrix $T_n$, however for 
our purposes the recursion relation itself is more useful, since it can
easily be implemented in an evolution program.
\item[e)] The inverse matrix $E_n^{-1}$ is also upper triangular with unit 
diagonal entries, and can be constructed noting that, for a generic 
nonsingular square matrix, the matrix of right eigenvectors (arranged in 
columns) is invertible, and the inverse is the matrix of left eigenvectors 
(arranged in rows). This leads to a recursion relation for the elements of 
$E_n^{-1}$, analogous to \eq{a3}. Setting $(E_n^{-1})_{i j} = \gamma_i^{(j)}$ 
we find
\beq
\gamma_i^{(j)} = \frac{1}{a_{i i} - a_{j j}} 
	\sum_{p = i}^{j - 1} \gamma_i^{(p)} a_{p j}~~,
\label{a4}
\eeq
\end{itemize}

It follows that triangular matrices can be diagonalized without computing 
any determinants. The implementation of the recursion relations 
Eqs.~(\ref{a3}-\ref{a4}) is in fact so efficient that the entire solution of 
the evolution equations (for reasonably sized matrices, say $n < 7-8$) can be 
performed analytically, for arbitrary $x_0$, and the chosen value of $x_0$ 
substituted only at the end.

\vskip 3.5cm
{\large {\bf Acknowledgements}}
\vskip 0.5cm

We would like to thank M. Caselle, G.~d'Agostini and P. Provero for 
discussions, T.~\c{C}uhadar, A.~Deshpande and E.~Rondio for informations 
on the fits of ref.~\cite{tul}, and R.~Ball for helping us to sharpen the
derivation of \eq{finsyst}.

\end{document}